\documentclass[prd,preprint,tightenlines,nofootinbib,eqsecnum,amsfonts,amssymb,amsmath]{revtex4-1}
\usepackage{graphicx}
\begin{document}
\newcommand{\dd}{{\rm d}}
\title{A two-mass expanding exact space-time solution}

\author{Jean-Philippe Uzan}
\email{uzan@iap.fr}
\affiliation{Institut d'Astrophysique de Paris, UMR-7095 du CNRS,
Universit\'e Paris-VI Pierre et Marie Curie, 98bis bd Arago, F-75014 Paris (France)}

\author{George F.R. Ellis}
\email{George.Ellis@uct.ac.za}
\affiliation{Astrophysics Cosmology and Gravitation Center,
Department of Mathematics and Applied Mathematics,
University of Cape Town, Rondebosch,
7701 Cape Town (South Africa)}

\author{Julien Larena}
\email{julien.larena@gmail.com}
\affiliation{Astrophysics Cosmology and Gravitation Center,
Department of Mathematics and Applied Mathematics,
University of Cape Town, Rondebosch,
7701 Cape Town (South Africa).\vskip2cm}

\begin{abstract}
\vskip0.25cm In order to understand how locally static
configurations around gravitationally bound bodies can be embedded
in an expanding universe, we investigate the solutions of general
relativity describing a space-time whose spatial sections have the
topology of a 3-sphere with two identical masses at the poles. We
show that Israel junction conditions imply that two spherically
symmetric static regions around the masses cannot be glued together.
If one is interested in an exterior solution, this prevents the
geometry around the masses to be of the Schwarzschild type and leads
to the introduction of a cosmological constant.  The study of the
extension of the Kottler space-time shows that there exists a
non-static solution consisting of two static regions surrounding the
masses that match a Kantowski-Sachs expanding region on the
cosmological horizon. The comparison with a Swiss-Cheese
construction is also discussed.
\end{abstract}

\date{\today}
\maketitle

\section{Staticity, spherical symmetry, and expansion}

A key issue as regards the nature of the universe is how the smooth
expanding large-scale universe is constructed from small-scale
extremely inhomogeneous domains that are essentially static. The
space-time in the Solar system for example is static to a very good
approximation, but it is very inhomogeneous ($\delta\rho/\rho \simeq
10^{+30}$). How can one construct an expanding universe solution by
gluing together large numbers of such quasi-static domains? The
static space-time regions representing local masses and the vacuum
regions around stars are of course structured into galaxies,
clusters of galaxies, and large scale structures such as walls and
voids; the same essential problem arises if we consider how they in
turn can be assembled into an expanding universe. For simplicity and
analytic clarity, we will assume that:
\begin{enumerate}
  \item the component massive entities are exactly spherically
symmetric;
  \item these entities, whether we think of them as stars, galaxies, or galaxy clusters,
  are embedded in an exact vacuum solution of the Einstein field equations
  (possibly with a non-zero cosmological constant, $\Lambda$);
  \item they are far enough apart from each other that the
vacuum space-time around each of them is also locally spherically
symmetric.
\end{enumerate}

Given the assumption of local spherical symmetry around each mass,
it is crucial that Birkhoff's theorem \cite{Bir23,HawEll73} will
apply:\\

4. the vacuum solution around each mass will necessarily be locally
static.\\

\noindent Thus we do not have to assume local staticity which
follows from the other previous assumptions. The local exterior space-time
(outside the massive entities) will thus locally be either a
Schwarzschild solution (if $\Lambda = 0$) or a Kottler solution (if
$\Lambda \neq 0$). So we can ask: how can one join exact
Schwarzschild or Kottler solutions together to form an expanding
universe?

\subsection{The Lindquist-Wheeler (L-W) solution}
This issue was tackled in a very innovative paper by Lindquist and
Wheeler a long time ago~\cite{LinWhe57}, using a Schwarzschild cell
method to model an expanding universe with closed spatial sections
(having the topology of a 3-sphere, $S^3$). For simplicity, they used a regular
lattice, which allows a very limited set of possibilities.
Considering $N$ vertices, they state that ``every vertex can be
equidistant from its neighbours only when $N = 5, 8, 16, 120,$ or
$600$", which corresponds to the most homogeneous
topologies of the 3-sphere~\cite{topoS3}. They then derived equations of motion for the expanding
universe from junction conditions between the cells. This work has
been recently revisited in an interesting way in Refs.~\cite{CliFer09,CliFer09a}.

This approach is quite different conceptually from the commonly used
``Swiss-cheese" construction~\cite{EinStr45,EinStr45a}. In that
case, one \emph{starts off} with a Friedmann-Lema\^{\i}tre spatially
homogeneous and isotropic geometry, and then cuts out spherical ``vacuoles"
within which individual masses are embedded. These masses are thus
contained in vacua within a spatially homogeneous fluid-filled
cosmos. In the Lindquist-Wheeler approach on the other hand, one starts off with the
inhomogeneous vacuoles alone, and then glues them together to construct
an emergent Friedmann-Lema\^{\i}tre geometry when averaged on large scales.
There is no fluid filling the space-time; rather (as in the case of
kinetic theory) fluid-like behaviour emerges on large scales when
one coarse-grains over the detailed structure. This is a far more
fundamental approach to the study of the relation between locally
static inhomogeneity and a globally expanding universe. However, this
method is not strictly self-consistent in that the gravitational
fields of the neighboring particles would in fact deform the field
in the neighborhood of each cell's vertex, thereby resulting in an
approximate rather than exact spherically symmetric space-time region
(the real solution will be locally a bit anisotropic about each
vertex). Nevertheless it is an acceptable approximation, and is a
very useful approach to tackle the issue raised here.

\subsection{Two-mass L-W solutions}

In order to understand the situation better, we investigate the
simplest possible such solution with isolated spherically symmetric
massive objects embedded in a vacuum with $S^3$ spatial topology:
namely the case with $N=2$, which Lindquist and Wheeler did not
consider~\cite{LinWhe57}. With the two masses at antipodal points of
the spatial sections, the space-time remains exactly spherically
symmetric about each mass - the approximation comment made above no
longer applies and an exact solution to the problem can be found.
The exact spherical symmetry around each mass ensures that
Birkhoff's theorem applies and the space-time is locally static.
Setting off radially from either one of the masses, in order to have
an $S^3$ spatial topology, the area of the surrounding 2-spheres
$S^2(r)$ at proper distance $r$ from the centre of symmetry must
reach a maximum and then decrease to zero at the antipodal point:
indeed we would expect to have two identical static solutions (if the 2 masses are
equal) back to back, joined at the equator of the $S^3$-spatial
section.

Two striking results now emerge (as will be shown in details below).
Firstly, if we want an exterior solution of this kind, we need a
positive cosmological constant; $\Lambda > 0$. Hence the two local
vacuum solutions to be joined together are Kottler space-times
\cite{Kot18,Wey19} rather than Schwarzschild solutions, filled in by
regular static bodies (e.g. stars) at their centers, so there is no horizon near the
centre of symmetry. Let us call this a Kottler exterior solution.

Secondly, we first show that Israel junction conditions
\cite{Isr66,Isr67} impose that the two space-times can only be joined on
an horizon or a surface of maximum area, where one of the metric
components vanishes. However, after constructing a coordinate system
regular on the horizon, we show that the matching cannot be
performed on the horizon. Gluing the two pieces together on a
surface of maximum area back to back would result in a static
configuration, but this is also shown to be impossible as no
coordinate system can regularly cover the neighbourhood of the
surface of maximum area. This prevents the existence of a static
space-time analog to the Einstein-Static 
model with the mass concentrated into two antipodal compact objects.
Instead, we find that the only solution contains two static regions
matched across their null horizons to a pair of Kantowski-Sachs
expanding and contracting universes; the expansion of the universe
is possible because of the existence of these spatially homogeneous
but anisotropic expanding regions. This solution is obtained by
considering the global structure of the exterior Kottler space-time,
which we show is essentially the same as that of the maximal de
Sitter hyperboloid \cite{HawEll73,Sch56}.

Surprisingly, this is again rather like the Swiss-Cheese models, but
now with the static domains embedded across null horizons (instead of space-like hypersurfaces
in the case of the Swiss-Cheese). Each
static and each expanding domain covers only part of the resulting
maximal Kottler exterior two-mass solution, in a way exactly
analogous to what occurs in the maximally extended de Sitter
solution. Furthermore, this shows that a globally static solution can exist
only at the price of introducing a (unphysical) surface layer.  A globally
non-singular 2-mass LW space-time does indeed exist, but it has
horizons separating spatially inhomogeneous static domains centred
on the two masses from spatially homogeneous time-evolving region.
It is this expanding universe domain that allows the two masses to
move away from each other, and so is the reason this universe model
can expand despite the static nature of space-time near each embedded
massive object.

The rest of the text will establish these results and is organized as follows: section II
is a quick summary of the statements of Birkhoff's theorem; section III will
present the results obtained when one tries to glue together two spherically
symmetric static space-times; section IV will discuss the results in a Swiss-cheese
configuration and comment on the analogy with the previous gluing approach;
finally section V will be a discussion of the results and their implications.

\section{Birkhoff's theorem}

Birkhoff's theorem stands for the crucial result that a vacuum spherically symmetric
space-time domain is necessarily either static or spatially
homogeneous. This is a local result, valid even if $\Lambda$ is
non-zero. In brief, the proof (see Ref.~\cite{Bir23} and also also Appendix B of
Ref.~\cite{HawEll73} and Refs.~\cite{Jeb05, Des05}) can be summarized by
considering a general spherically symmetric space-time with metric
\begin{equation}
 \dd s^2 =-A(t,\chi)\dd t^2 + B(t,\chi)\dd\chi^2 + \chi^2\dd\Omega^2,
\end{equation}
(without loss of generality, we choose $\chi$ as an area coordinate).
First, as long as the stress-energy tensor components of the
source satisfy $T_{t\chi}=0$, the Einstein equations imply that
$\partial_t B=0$. This implies that $B = B(\chi)$, which is the
first part of the Birkhoff theorem.
Then, the combination $G^t_t-G^\chi_\chi=0$, that holds for a vacuum as well as a
cosmological constant, implies that $B' A + B A' =0$.
It follows that $A$ can be chosen so that $A = A(\chi)$ and
$A(\chi)B(\chi)=K=1$, so that
\begin{equation}\label{eq1}
 \dd s^2 =-A(\chi)\dd t^2 + A^{-1}(\chi)\dd\chi^2 +
 \chi^2 \dd\Omega^2.
\end{equation}
The metric is static if $A>0$ (surfaces of constant $t$ are
spacelike) and spatially homogeneous if $A<0$ (surfaces of constant
$t$ are timelike).

Thus, the existence of an extra symmetry for the metric derives from
its spherical symmetric nature and the fact that
$T_{t\chi}=0$ and $T^t_t-T^\chi_\chi=0$, as is valid in the exterior region
surrounding a point mass or an interior static solution (such as the
Schwarzschild interior solution). In the latter case, which is the
one of interest to us, by continuity the exterior region close to
the static spherically symmetric object represented by the solution
is chosen to be static rather than spatially homogeneous.

\section{Gluing spherically-symmetric space-times}

\subsection{Case of a spherically-symmetric static boundary}

\begin{figure}[htp]
\vskip-.5cm
\includegraphics[width=12cm]{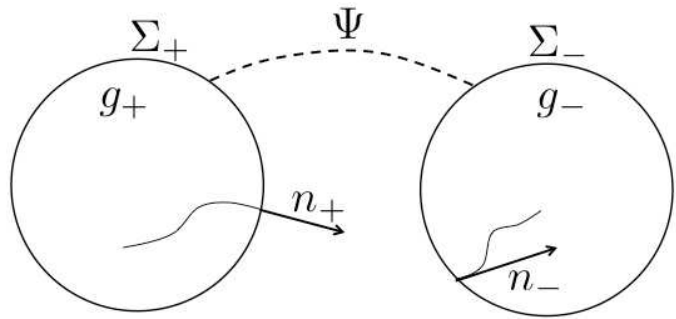}
\vskip-3cm
\caption{Geometry of the matching between two space-times with
metrics $g_\pm$ and
boundaries $\Sigma_\pm$ that are identified via the mapping $\Psi$.
$n_\pm$ are the vectors normal to the hypersurfaces $\Sigma_\pm$.}
\label{fig00}
\end{figure}

Let us consider two space-times with metric~(\ref{eq1}) and coordinate systems $(t_\pm,r_\pm,\theta,\varphi)$ in
each of the two regions so that the metrics take the form
\begin{equation}\label{eq1b}
 \dd s_\pm^2 =-A(r_\pm)\dd t_\pm^2 + A^{-1}(r_\pm)\dd r_\pm^2 +
 r_\pm^2\dd\Omega^2.
\end{equation}
We glue these space-times together on spacelike hypersurfaces defined by
$\Sigma_\pm=\lbrace r_\pm = R_\pm\rbrace$ with $R_\pm$ constants,
assuming we identify the angular coordinates $\theta_\pm=\theta$
and $\varphi_\pm=\varphi$ (see Figure \ref{fig00}). The two normal vectors,
chosen so that they are continuous across the junction ($n^{(+)}_\mu$
points out of its domain and  $n^{(-)}_\mu$ into its domain), are
given by $n^{(\pm)}_\mu = \pm A_\pm^{-1/2}\delta_\mu^{r_\pm}$. The
induced metrics and extrinsic curvatures are thus given by
\begin{eqnarray}
 \gamma^{(\pm)}_{ab}\dd x^a \dd x^b = -A_\pm\dd t^2_\pm +
 R^2_\pm\dd\Omega^2,\label{inducedg1}\\
K_{ab}^{(\pm)}dx^adx^b=\pm\sqrt{A_{+}}\left(-\frac{1}{2}A'_{\pm}\dd t_{\pm}^2+
 R_{\pm}d\Omega^{2}\right).\label{extrinsicK1}
\end{eqnarray}

It follows from Eq.~(\ref{inducedg1}), using the standard Israel
junction conditions~\cite{Isr66,Isr67,GolKat95}, that the continuity
of the induced metric imposes that $R_+=R_-\equiv R$ and that
$A_+(R)\dd t_+^2 = A_-(R)\dd t_-^2$. It then follows from
Eq.~(\ref{extrinsicK1}) that the extrinsic curvature can be
continuous only if $A_\pm(R)=0$, which occurs at a horizon. In
conclusion, {\it only on a horizon
($A = 0$) may two spherically symmetric static space-times possibly
be glued back to back.}

We cannot use the Schwarzschild coordinates to perform the matching
at the horizon because this coordinates system is singular there.
Let us assume that $\hat r$ is such that $A(\hat r)=0$ and let us Taylor
expand $A$ in a neighborhood of $\hat r$ as $A(r)=\lambda\times
(r-\hat r)$ with $\lambda\equiv A'(\hat r)\not=0$. Using the
coordinate $r_*$ defined by $\dd r_*/\dd r=A^{-1}(r)$ we deduce that
$r_*\simeq \lambda^{-1}\log (r-\hat r)$ in a neighborhood of $\hat
r$. Introducing the null-coordinates $u=t+r_*$ and $v=t-r_*$ and the
relabelling $U=\exp(\lambda u/2)$ and $V=\exp(-\lambda v/2)$, the
metric in the neighborhood of $\hat r$ takes the form $\dd s^2=
\frac{4}{\lambda}\dd U\dd V+r^2\dd\Omega^2$, and is regular in
$r=\hat r$, that is on $UV=\exp(\lambda r_*)$. $r$ is defined in
such a way that $4\pi r^2$ is the surface area of a 2-sphere of
radius $r$, which defines it uniquely.

Now, we can study the matching between two space-times
\begin{equation}
 \dd s_\pm^2= \frac{4}{\lambda_\pm}\dd U_\pm\dd V_\pm+r_\pm^2\dd\Omega^2.
\end{equation}
Any spherically symmetric spacelike hypersurface  $\lbrace r_\pm=
R_\pm\rbrace$ corresponds to a hypersurface of equation $\lbrace
U_\pm V_\pm={\rm const.}\rbrace$ so that the normal unit spacelike
vector is
$$
n_\mu^{(\pm)}=\frac{\pm1}{\sqrt{\lambda_\pm U_\pm V_\pm}}\left(V_\pm\delta_\mu^{U_\pm}
+ U_\pm\delta_\mu^{V_\pm} \right).
$$
The induced metric is then given, as previously, by
$$
 \gamma^{(\pm)}_{ab}\dd x^a \dd x^b = -\frac{1}{\lambda_\pm U_\pm V_\pm}
 \left(V_\pm\dd U_\pm - U_\pm\dd V_\pm\right)^2
 +R_\pm^2\dd\Omega^2
$$
so that its continuity implies, as previously, $R_+=R_-\equiv R$. Now,
the continuity of the extrinsic curvature implies that
$K_{\theta\theta}^{(\pm)}=-\Gamma_{\theta\theta}^\alpha
n_\alpha^{(\pm)}$ is continuous, which is impossible. This show that
{\em the back to back matching of two static solutions cannot be
performed on the horizon $A(r) =0$}.

Thus, it is {\it impossible} to glue two spherically symmetric static space-times on a spherically
symmetric and static boundary, even if the boundary is a horizon.

\subsection{Case of a dynamical boundary}

In the previous argument, we required the gluing surface to be a constant $r$ hypersurface;
in order to check if this is a restrictive assumption, let us generalize to the situation in which the
boundary is moving, i.e. $R_\pm = R_\pm(t_\pm)$.  The two matching hypersurfaces are
defined by $\Sigma_\pm=\lbrace r_\pm - R_\pm
(t_\pm)={\rm const}\rbrace$, that we assume to be spacelike (we
separately show that the null-surface case is indeed possible, see
below). The two normal vectors are then given by
$$
n^{(\pm)}_\mu = \frac{\pm1}{\sqrt{A_\pm - \dot R_\pm^2/A_\pm}}(-\dot R_\pm\delta_\mu^{t_\pm}+\delta_\mu^{r_\pm}),
$$
where $\dot R_\pm=\dd R_\pm/\dd t_\pm$. The induced metrics and extrinsic curvatures are given by
\begin{equation}
 \gamma^{(\pm)}_{ab}\dd x^a \dd x^b = - \frac{A_\pm^2-\dot R_\pm^2}{A_\pm} \dd t^2_\pm + R^2_\pm\dd\Omega^2
\end{equation}
and
\begin{equation}
 K^{(\pm)}_{ab}\dd x^a \dd x^b =  \frac{\pm1}{A_\pm\sqrt{A_\pm-\frac{\dot R_\pm^2}{A_\pm}}}
 \left[\frac{1}{2}\left(3\dot R_\pm^2A'_\pm-2\ddot R_\pm A_\pm -A^2_\pm A'_\pm \right)\dd t_\pm^2
 + R_\pm A_\pm^2\dd\Omega^2
 \right]
\end{equation}
using that on $\Sigma_\pm$, $\dd r_\pm=\dot R_\pm\dd t_\pm$. As
$R_\pm$ and $A_\pm$ are both positive functions, it is impossible
to have continuity of $K_{\theta\theta}$ unless
$A_\pm[R_\pm(t_\pm)]=0$ for all $t_\pm$. This imposes that $\dot R_\pm=0$, and thus,
a configuration with $\dot R_\pm\not=0$ cannot be constructed if
the join surfaces are timelike.

\subsection{Conclusions}

To summarize, two spherically symmetric static solutions cannot be glued together back-to-back to
obtain a static space-time with the spatial topology of a 3-sphere. In particular,
this implies that we cannot glue two Schwarzschild exterior solutions together
to get a 2-mass space-time with $S^3$ topology.
This can be possible only at the expense of introducing a surface layer, which we
do not want to consider.

\section{Two mass solution}

We now show that a spatially closed space-time with two masses the spatial size of which
is larger than their Schwarzschild horizon can be constructed using two copies of the exterior
Kottler space-time~ \cite{Kot18,Wey19}, which is the generalisation of the Schwarzschild solution
to incorporate a non-vanishing cosmological constant. This solution has two horizons, a black-hole horizon at
$r=r_b$ similar to the Schwarzschild horizon and a cosmological horizon at $r=r_c$ similar to the de Sitter horizon.

As we shall see, a solution with a $S^3$ spatial topology with two
antipodal masses can be constructed but with an expanding domain in
between the two static domains, as will become clear once we
construct the Penrose-Carter diagram for the solution.

\subsection{Kottler space-time}

The Kottler solution \cite{Kot18,Wey19} (for a summary see
Ref.~\cite{Per04}) is the extension of the Schwarzschild solution to
include a cosmological constant,
\begin{eqnarray}\label{eq1c}
 \dd s^2 &=&-A(r)\dd t^2 + A^{-1}(r)\dd r^2 + r^2\dd\Omega^2,\\
 A(r) &=& 1 - \frac{2GM}{r} - \frac{\Lambda r^2}{3}.
\end{eqnarray}
It is easy to check that the Killing vector
\begin{equation}
 \xi^\mu = \delta^\mu_0
\end{equation}
has norm $g_{\mu\nu}\xi^\mu\xi^\nu=A(r)$ and is thus timelike as
long as $A>0$. We thus have two cases:
\begin{itemize}
\item If $9(GM)^2\Lambda>1$, $A$ is negative for $r>0$ so that $ \xi^\mu$ is spacelike
and the space-time contains no static region but is spatially
homogeneous. We exclude this case as it does not allow a static
central body.
\item If $9(GM)^2\Lambda<1$, $A$ is positive for $r$ between $r_b$ and $r_c>r_b$ which correspond respectively
to the black-hole and cosmological horizons. They can be found to be
given by
\begin{equation}
 r_c =\frac{2}{\sqrt{\Lambda}}\cos\left(\frac{\psi}{3} + \frac{\pi}{3} \right),
 \qquad
 r_b =\frac{2}{\sqrt{\Lambda}}\cos\left(\frac{\psi}{3} - \frac{\pi}{3} \right),
 \qquad
 \cos\psi = 3GM\sqrt{\Lambda}
\end{equation}
so that we have
$$
 2GM<r_b<3GM<\frac{1}{\sqrt{\Lambda}}<r_c<\frac{3}{\sqrt{\Lambda}}.
$$
The space-time is thus static in the region $r_b<r<r_c$. It is clear
that the third root is negative since
$A=-(\Lambda/3r)(r-r_b)(r-r_c)(r+r_b+r_c)$. $r=r_b$ and $r=r_c$ are
Killing horizons since $\xi$ vanishes on these hypersurfaces. One
can check that for both values, $A'(r)\not=0$ (the horizons are not
degenerate) and $A''(r)<0$.
\end{itemize}
We consider solutions of this second type where an interior
solution representing a spherically symmetric body occupies the
region $0< r < R_M$. This requires that $R_M > r_b$, so the
spatially homogeneous vacuum region for $0 < r < r_b$ is filled in
by the body; hence there is no horizon at $r = r_b$.
We require that $R_M< r_c$ (there is some vacuum region around the
central body). Such a solution will be called a \emph{Kottler exterior solution}.

The metric can be conveniently rewritten in terms of the radial coordinate $r_*$
defined by
\begin{equation}
 \dd r_* \equiv\frac{\dd r}{A(r)}
\end{equation}
that is explicitly given by
\begin{equation}\label{rstareq}
 r_* = \frac{3}{\Lambda}\left[
 \frac{r_b\ln(r-r_b)}{(r_c-r_b)(r_b-r_s)} -
 \frac{r_c\ln(r-r_c) }{(r_c-r_b)(r_c-r_s)} +
 \frac{r_s\ln(r-r_s)}{(r_b-r_s)(r_c-r_s)}\right]
\end{equation}
with $r_s\equiv-(r_b+r_c)=-2/\sqrt{\Lambda}\cos(\psi/3)$.

\subsection{Penrose-Carter diagram}

To explicitly describe this solution, let us construct its
Penrose-Carter diagram (see Ref.~\cite{HawEll73}). Obviously, the
metric can be rewritten using the ingoing Eddington-Finkelstein
coordinates as
\begin{equation}
 \dd s^2 =-A \dd u^2 + 2\dd u\dd r + r^2\dd\Omega^2,
\end{equation}
with $\dd u=\dd t+\frac{1}{A}\dd r=\dd t+\dd r_*$ or the outgoing Eddington-Finkelstein coordinates as
\begin{equation}
 \dd s^2 =-A \dd v^2 - 2\dd v\dd r + r^2\dd\Omega^2,
\end{equation}
with $\dd v=\dd t-\frac{1}{A}\dd r=\dd t-\dd r_*$.
It is clear under the first form that
$\partial_u$ is a Killing vector \cite{Boy69}. When $A>0$, it is spacelike and one recovers
the Kottler solution under its static form (set $r=v$ and $\dd t=\dd u - \dd v/A$)
while when $A<0$ the Killing vector is timelike and one gets a homogeneous solution
(set $t=v$ and $\dd r=\dd u - \dd v/A$). Using the null-coordinates $(u,v)$ leads to the usual form
\begin{equation}
 \dd s^2 =-A \dd u\dd v + r^2\dd\Omega^2,
\end{equation}
which remains pathological when $A=0$. As usual with horizon, the surfaces of
constant $u$ or constant $v$ (resp. ingoing/outgoing null geodesics) are geometrically
well defined but their labelling is not on the horizon.

When focusing on the exterior Kottler solution, the only horizon of interest is located in $r=r_c$ (remind
that $r_s<0<r_b<R_M<r_c$).  Introducing
$$
 U = \exp\left[-\alpha \frac{u}{2}\right],
 \qquad
 V = \exp\left[ \alpha \frac{v}{2}\right],
$$
with $\alpha=(\Lambda/3)(r_c-r_b)(r_c-r_s)/r_c$, the metric becomes
\begin{equation}
 \dd s^2 =-B \dd U\dd V + r^2\dd\Omega^2,
\end{equation}
with $B\equiv -(4A/\alpha^2)\exp(\alpha r_*)$. Given the expression~(\ref{rstareq}),
it is explicitly given
$$
 B = \frac{\Lambda}{3r}\frac{4}{\alpha^2}
  (r-r_b)^{1+\frac{r_b(r_c-r_s)}{r_c(r_b-r_s)}}
  (r-r_s)^{1+\frac{r_s(r_c-r_b)}{r_c(r_b-r_s)}},
$$
which does not vanish for $r\in [R_M,+\infty [$. This is similar to the standard analysis to
construct a coordinate system regular on the horizon of a Schwarzschild black-hole (see
e.g. Ref.~\cite{MTW}).
We refer to Refs.~\cite{lake} for the definition of a maximally extended map that covers
the Kottler solutions and that generalizes the maps introduced in Refs.~\cite{map1,map2}
for Schwarzschild and Reissner-Nordstr\"om.
For a detailed description of the global structure and horizons of the Kottler space-time see
Refs.~\cite{LakRoe77,Stu83}.

\begin{figure}
\includegraphics[width=10cm]{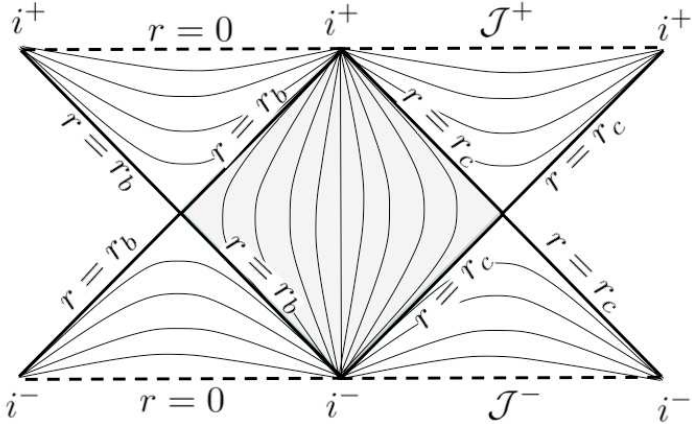}
\caption{Penrose-Carter diagram of a Kottler space-time.
Thin lines represent constant $r$ hypersurfaces. The space-time is
static in the central lozenge region and homogeneous in the other
regions.} \label{fig0}
\end{figure}

\begin{figure}
\includegraphics[width=10cm]{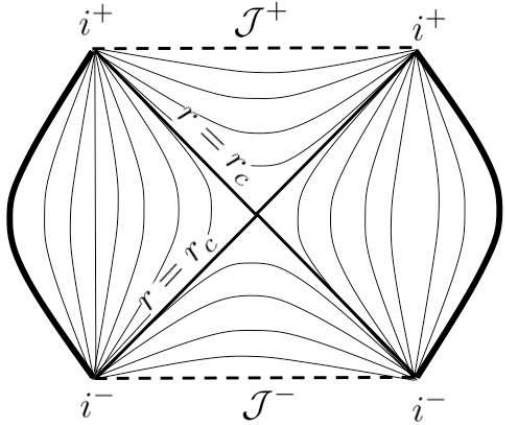}
\caption{Exterior two-mass Kottler solution.}
\label{fig2}
\end{figure}

From this latter form, we can deduce the Penrose-Carter diagram
depicted on Fig.~\ref{fig0}. The Penrose-Carter diagram of our
exterior solution with two identical masses is only a subset of the
general diagram and is depicted on Fig.~\ref{fig2}. It is obtained
by gluing another copy of the space-time described
on the right-hand side of figure~\ref{fig0} and by
cutting the two static regions at a given radius $R_{M}>r_{b}$
(represented by the thick lines) to restrict to the exterior
solution of interest. We see that it exhibits two static regions
around each mass extending up to the cosmological horizons (at
$r=r_c$). The two horizons cannot be glued together and there must
exist a homogeneous region in between them. In this homogeneous
region connecting the two static regions, the $r$ coordinate is
effectively a timelike coordinate whereas $t$ is a radial
coordinate. Note that the two massive bodies are causally
disconnected: light emitted in the static region surrounding one of
the masses will eventually cross the first horizon, reach the
expanding region and attain $\mathcal{J}^+$ (that is timelike) but
will never cross the second horizon. Note also that the Killing
vector cannot be timelike everywhere, since the horizon is a null
surface. This is the reason why there is no analog of the Einstein
static Universe with the masses concentrated in two antipodal
compact objects. Let us also note that similar diagrams have
been constructed in Ref.~\cite{klb} while studying constant tension stars
and hybrid stars (i.e. having an interior zone with negative pressure
and an infinitely thin outermost layer with positive pressure and
energy density) in a Schwarzschild spacetime.

\subsection{Embedding}

This general construction can be understood by an embedding in a
5-dimensional Minkowski space-time \cite{Mar99,GibMarGar04}. Indeed, the exact form of the
embedding cannot be determined analytically but assuming the
masses are such that $r_b\ll r_c$, the structure of the space-times
on cosmological scales are closed to the one of a de Sitter
space-time. Nevertheless, the worldlines of the two masses do not have the symmetries of a de Sitter space-time and therefore induces a preferred slicing of space-time; hence resulting in a preferred time direction.

The expanding region has a metric given by
\begin{equation}
 \dd s^2 = -\frac{1}{A(t)}\dd t^2 + A(t)\dd r^2 + t^2\dd\Omega^2.
\end{equation}
Defining $\dd\tau =\dd t/\sqrt{A(t)}$ and $B^2(\tau) =
A(t(\tau))$, it takes the form
\begin{equation}
 \dd s^2 = -\dd \tau^2 + B^2(\tau)\dd r^2 + t^2(\tau)\dd\Omega^2.
\end{equation}
This anisotropic expanding region is the interior solution of a
black-hole (see e.g. \cite{DorLobCra06}) in the vacuum case, but
is of the Kantowski-Sachs form when filled with a fluid (and
possibly a cosmological constant) \cite{KanSac66}. This is thus another kind of
Swiss cheese model but within a anisotropic vacuum space-time in
the expanding region.

\begin{figure}[htp]
\vskip-.25cm
\includegraphics[width=12cm]{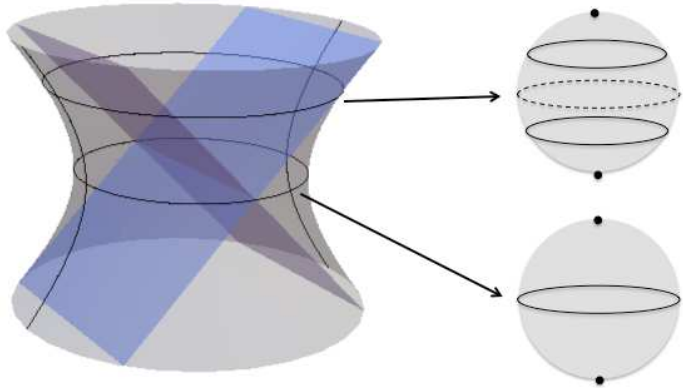}
\vskip-1cm
\caption{Embedding of the solution in a 5-dimensional space-time. The hyperboloid
is a 4-dimensional space-time. Each circle is a spacelike section which has the topology
of the 3-sphere. The two lines represent the worldlines of the masses. For each slice, the space has the
topology of a 3-sphere with a mass at the North pole and a mass at the South pole.
We have depicted the horizon (solid line).}
\label{fig3}
\end{figure}

Taking horizontal sections at different times, we get $S^3$ spatial
slicings that contain back to back copies of a massive object
surrounded by a static vacuum that is in turn surrounded by an
expanding universe domain that has a cross sectional area that
increases to a maximum, where it is matched to an identical solution
in back to back fashion (see the inset diagram in Fig 4). As time
progresses, the size of the expanding universe section decreases
from most of the spatial section (at large negative times) to zero
and then increases again to most of the spatial sections (at large
positive times). This is like the Wheeler analysis of the time
evolution of the throat in a black hole universe.

\subsection{Conclusion}

This gives a two-mass version of the de Sitter expanding universe. It can be
envisioned as two world lines added antipodally to the maximal
$S^3\times R$ de Sitter hyperboloid. If these are of very small
mass, by continuity the global structure \cite{HawEll73,Sch56} will
remain unchanged. Indeed this is the structure of the two-mass exterior Kottler
solution, as can be seen by filling in the two Schwarzschild
horizons in the maximal Kottler space-time.

\section{The Swiss-Cheese approach}

A standard method to embed a single mass in an expanding space-time
is to use the Einstein-Strauss method \cite{EinStr45,Kan69} to
construct a Swiss-cheese model \cite{BalBerCom88}. We thus consider
the two space-times
\begin{equation}\label{eqint}
 \dd s^2 =-A(r)\dd t^2 + \frac{\dd r^2}{A(r)} + r^2\dd\Omega^2,
\end{equation}
and
\begin{equation}\label{eqext}
 \dd s^2 =-\dd T^2 + a^2(T)\left(\dd\chi^2 + \sin^2\chi\dd\Omega^2\right),
\end{equation}
describing the geometry of a Kottler space-time and of a
Friedmann-Lema\^{\i}tre (FL) universe with spherical spatial sections
having chosen units such that the comoving curvature radius $R_c=1$. We
decide to glue these two space-times on a constant $r=r_0(t)$
hypersurface in the Kottler space-time and on a  $\chi=\chi_*$
hypersurface. In the FL region,  the normal is
given by  $n^{({\rm FL})}_\mu=\delta_\mu^\chi/a$ so that
\begin{equation}
 \gamma^{({\rm FL})}_{ab}\dd x^a \dd x^b = - \dd T^2 + a^2(T)\sin^2\chi_*\dd\Omega^2
 ,\qquad
 K^{({\rm FL})}_{ab}\dd x^a \dd x^b = -a(T)\sin\chi_*\cos\chi_*\dd\Omega^2.
\end{equation}
We stress that we note $\dot a = \dd a/\dd T$ and $\dot r_0=\dd r_0/\dd t$.
Since
\begin{equation}
 \gamma_{ab}^{({\rm K})}\dd x^a \dd x^b = - \frac{A^2-\dot r_0^2}{A} \dd t^2 + r_0^2\dd\Omega^2,
\end{equation}
the continuity of the induced metric implies that
\begin{equation}
r_0(t) = a(T)\sin\chi_*,\qquad
\frac{\dd T}{\dd t} = \sqrt{\frac{A^2[r_0(t)]-\dot r_0^2(t)}{A[r_0(t)]}},
\end{equation}
which defines the worldsheet of the hypersurface on which we match in the Kottler region
and the relation between the times in both regions.

Now in the FL region, the scale factor must satisfy the Friedmann
equation
\begin{equation}
 \left(\frac{\dot a}{a}\right)^2 = \frac{8\pi G}{3}\rho - \frac{1}{a^2} + \frac{\Lambda}{3},
\end{equation}
$\rho=\rho_0(a_0/a)^3$ being the energy density of a pressureless fluid.
The continuity of the extrinsic curvature is achieved only if the cosmological constant
is the same in the two space-times and if
\begin{equation}
 M = \frac{4\pi}{3} \rho a^3 \sin^3\chi_*.
\end{equation}
It follows that
\begin{equation}\label{tofT}
\left(\frac{\dd T}{\dd t}\right)^2\cos^2\chi_* = A^2[a(T)\sin\chi_*].
\end{equation}
Hence given a FL space-time with pressureless matter and a cosmological constant, we know $a(T)$
and we can insert a spherical region of radius $\sin\chi_*$ which contains a constant
mass $M =(4\pi\rho a^3 \sin^3\chi_*)/3$ at its center and has the Kottler geometry. Then Eq.~(\ref{tofT}) gives
the relation between the coordinate times of the two space-times.

The limit $\chi_*\rightarrow \pi/2$ is such that $A\rightarrow 0$, i.e. the mass becomes such that
the equatorial 2-sphere is also the cosmological horizon of the Kottler space-time.

\begin{figure}[htp]
\vskip-.25cm
\includegraphics[width=9.5cm]{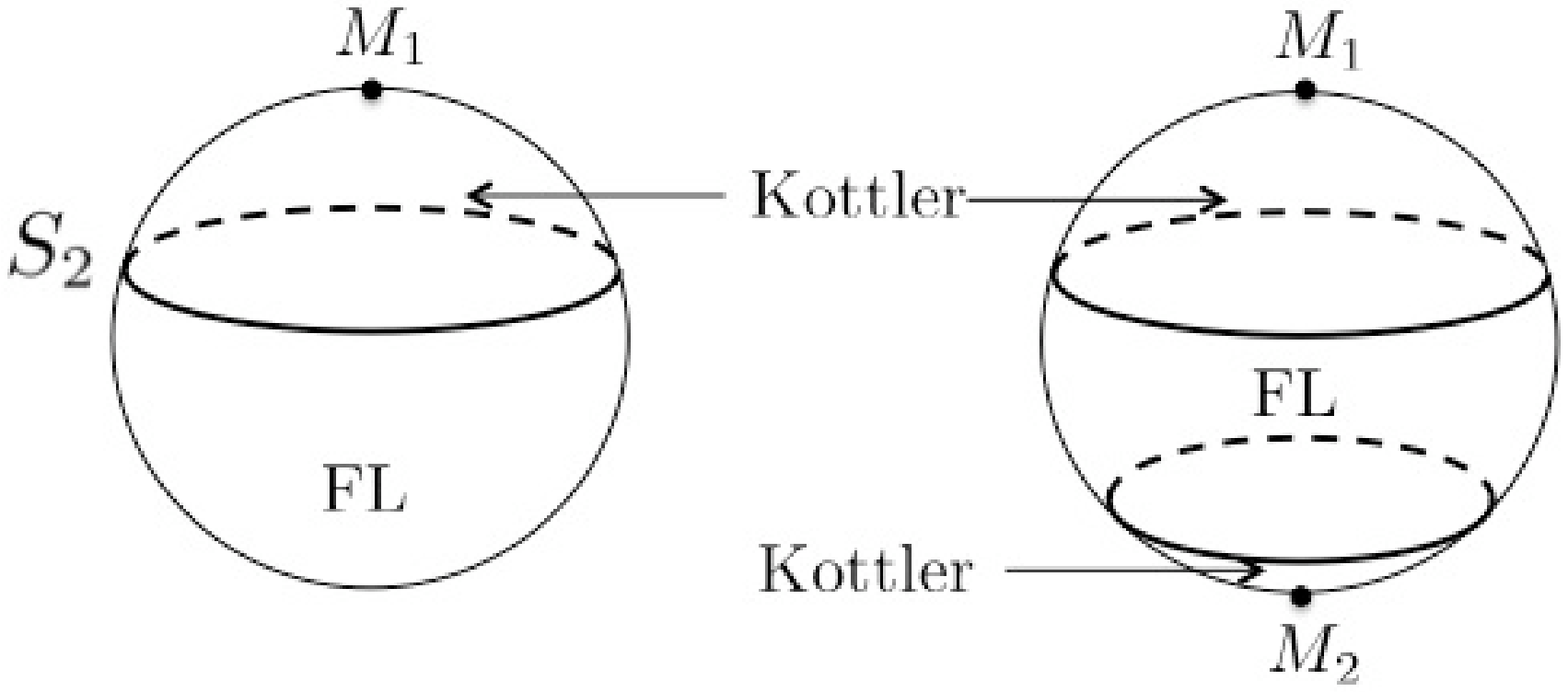}\hskip-2cm\includegraphics[width=8.5cm]{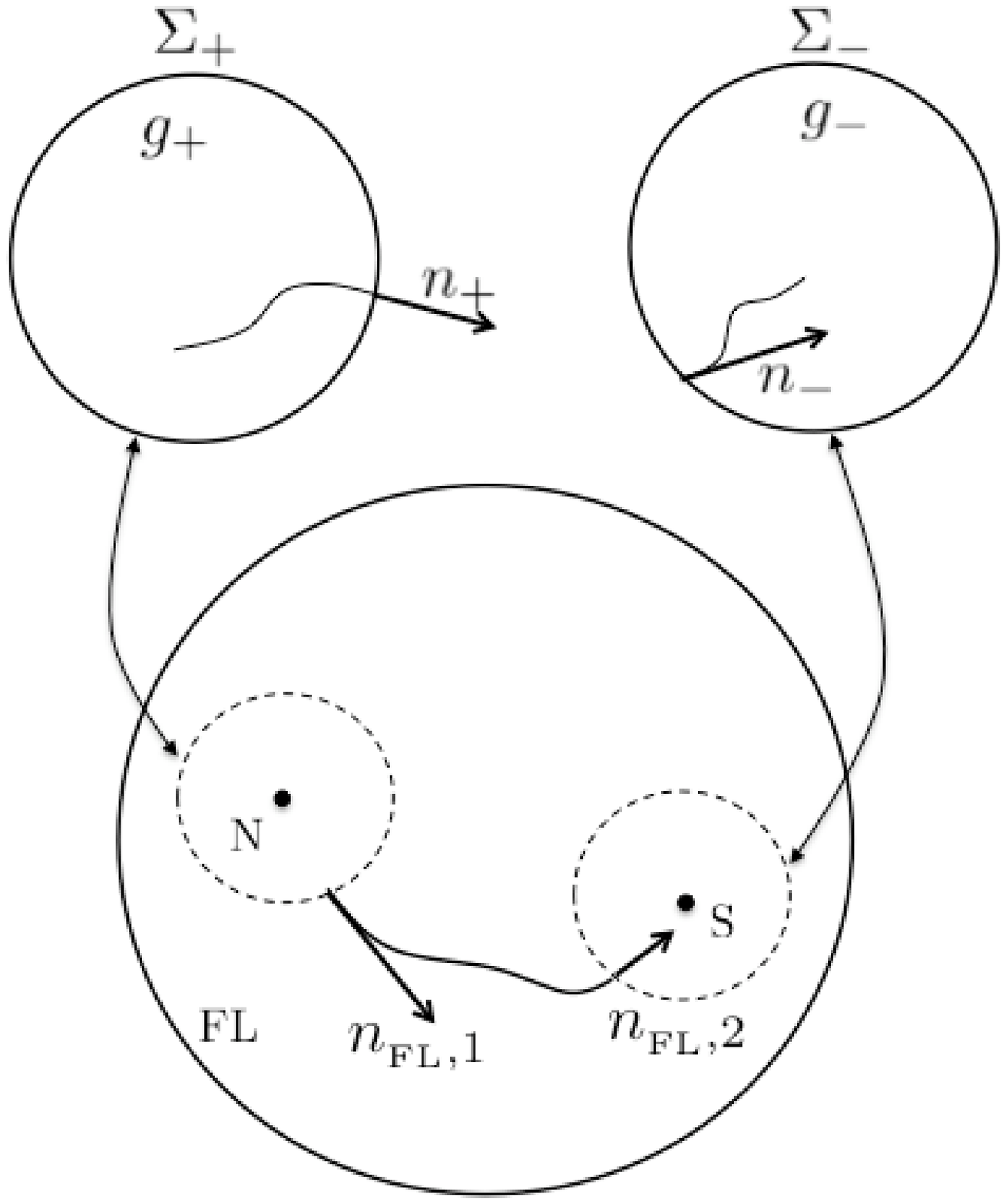}
\vskip-.5cm
\caption{ (Left) Construction of two Kottler holes in a FL space-time with spherical spatial sections.
The mass inside each hole depends on the size on the hole. (Right) Consider two Kottler
holes (respectively with boundaries $\Sigma_+$ and $\Sigma_-$ and associated normal
vectors $n_+$ and $n_-$) inside a FL spacetime (with $n_{\rm FL1}$ and $n_{\rm FL2}$ for
the normal vectors associated with the two boundaries). The two holes can be
extended in order to join smoothly only if their centers are antipodal. In that limit
$n_{\rm FL1}\rightarrow n_{\rm FL2}$. The two holes thus contain the
same mass but there remains a surface layer related to the jump of the extrinsic curvature.}
\label{fig4}
\end{figure}

With such a solution, we can indeed insert two Kottler holes in the FL solution and make their size increase
(see Fig.~\ref{fig4}-left). Indeed, if they are not antipodal then the two boundaries (which are 2-sphere)
will not match. If they are, we can increase their size until they join, at which point they will
contain, by construction the mass. The induce metric on the boundary will be continuous but,
as can be seen from Fig.~\ref{fig4}-right, the extrinsic curvature will not be continuous
(in particular because the unit normal vector is going in the FL spacetime for one hole
and outside for the other). When the two holes join, they are glued on their horizon and
we are back to the situation of Fig.~\ref{fig00}. We thus recover the solution of the first section.

\section{Discussion}

The LW models are significant tools in elucidating the ongoing
discussion on averaging in cosmology, and the relation between small
scale inhomogeneous models and large scale spatially homogeneous
models (see e.g. \cite{CliFer09, CliFer09a,Wil09, Wil09a,
Maretal07}). This could provide a better understanding of the
emergence of a fluid-like description of the Universe on large
scales from a more realistic local clumpy distribution of matter. In
the line of this program, we have analyzed the geometrical structure
of an exact two-mass singularity free version of the LW models: an
expanding model which is the analogue of the maximal de Sitter
hyperboloid once it has been embedded in a 5 dimensional space-time.
This resolves the paradox raised in the introduction to this paper:
how can locally static domains be glued together to give an
expanding universe? Indeed, the solution presented here is locally
static around each compact object but not globally so even if the
conditions of the Birkhoff theorem still holds (spherical symmetry
around each mass and vacuum + cosmological constant solution of
Einstein equations). This is different from a de Sitter space-time,
which is locally static everywhere: nothing intrinsic to the de
Sitter space-time identifies any particular null surface as the
horizon, and there are local timelike and null Killing vector fields
at every point. The 2-mass exterior Kottler solution, on the other
hand, is not locally static in the region across the horizon: as in
the case of the Schwarzschild solution (inside its horizon), there
is no timelike Killing vector field there, and the horizon is
uniquely defined locally by existence of a null Killing vector
field.

We have also shown that it is not possible to construct a globally
static solution that is the analogue of the Einstein Static universe
(a timelike Killing vector field occurs at every point), but with
the mass concentrated in two antipodal massive bodies. This would
require the introduction of an unphysical surface layer. We have
shown that the expanding region can either be filled with a
cosmological constant and has a Kantowski-Sachs geometry, or by a
pressureless fluid and a cosmological constant. In that latter case,
the geometry of the expanding region can be of the FL type.\\

\noindent{\bf Acknowledgements:} We thank Timothy Clifton for
his fruitful comments on an early version of this text. We also thank
Chris Clarkson, Gilles Esposito-Far\`ese, Charles Hellaby, Jeff Murugan, Tony
Rothman and Amanda Weltman for discussions. JPU thanks UCT
for hospitality during the period this project was initiated. JL is supported by the Claude Leon Foundation (South Africa).

\end{document}